\documentclass[12pt,eqsecnum,preprint]{aastex}

\begin{document}

\title{The Physical Nature of Polar Broad Absorption Line Quasars}
\author{Kajal K. Ghosh\altaffilmark{1}} \and\author{Brian
Punsly\altaffilmark{2}} \altaffiltext{1}{Universities Space Research
Association, NASA Marshall Space Flight Center, VP62, Huntsville,
AL, USA} \altaffiltext{2}{4014 Emerald Street No.116, Torrance CA,
USA 90503 and International Center for Relativistic Astrophysics,
I.C.R.A.,University of Rome La Sapienza, I-00185 Roma, Italy,
brian.m.punsly@L-3com.com or brian.punsly@gte.net}

\begin{abstract}
It has been shown based on radio variability arguments that some
BALQSOs (broad absorption line quasars) are viewed along the polar
axis (orthogonal to accretion disk) in the recent article of Zhou et
al. These arguments are based on the brightness temperature, $T_{b}$
exceeding $10^{12}\,^{\circ}$ K which leads to the well-known
inverse Compton catastrophe unless the radio jet is relativistic and
is viewed along its axis. In this letter, we expand the Zhou et al
sample of polar BALQSOs to the entire SDSS DR5. In the process, we
clarify a mistake in their calculation of brightness temperature.
The expanded sample of high $T_{b}$ BALQSOS, has an inordinately
large fraction of LoBALQSOs (low ionization BALQSOs). We consider
this an important clue to understanding the nature of the polar
BALQSOs. This is expected in the polar BALQSO analytical/numerical
models of Punsly in which LoBALQSOs occur when the line of sight is
very close to the polar axis, where the outflow density is the
highest.
\end{abstract}

\keywords{(galaxies:) quasars: absorption lines --- galaxies: jets
--- (galaxies:) quasars: general --- accretion, accretion disks --- black hole physics}

\section{Introduction}About 15\% - 20\% of quasars show broad UV absorption lines (loosely
defined as absorbing gas that is blue shifted at least 5,000 km/s
relative to the QSO rest frame and displaying a spread in velocity
of at least 2,000 km/s) \citep{wey97,hew03,rei03}. The implication
is that large volumes of gas are expelled as a byproduct of
accretion onto the the central supermassive black hole.
Understanding the details of this ejection mechanism is a crucial
step towards understanding the physics of the central engine of
quasars. Although evolutionary processes might be related to BAL
outflows, it is widely believed that all radio quiet quasars have
BAL flows, but the designation of a quasar as a BALQSO depends on
whether the line of sight intersects the solid angle subtended by
the outflow. The standard model of quasars is one of a hot accretion
flow onto a black hole and a surrounding torus of molecular gas
\citep{ant93}. The BAL outflow can be an equatorial wind driven from
the luminous disk that is viewed at low latitudes, just above the
molecular gas, \citet{mur95}, or a bipolar flow launched from the
inner regions of the accretion flow \citep{pun99,pun00}. The most
general feature of 3-D simulations of accretion flows onto black
holes is huge ejections of gas from the central vortex of the
accretion flow along the polar axis which rivals the accreted mass
flux for rapidly spinning black holes \citep{dev05,haw06}. Thus, it
is of profound theoretical interest to look for evidence of these
endemic polar ejecta.
\par BALQSOs are so distant that direct imaging of the BAL
region is beyond the resolution of current optical telescopes. Thus,
much of the discussion of BAL geometry is based on deductive
reasoning. A novel idea for finding the orientation of BALQSOs was
developed in \citet{zho06}. They used radio variability information
to bound the size of the the radio emitting gas then deduced that
the radio emission must be viewed close to the polar axis and
emanate from a relativistic jet, thereby avoiding the well known
inverse Compton catastrophe. They found BALQSOs that satisfy these
conditions in SDSS DR3. We expand their methods to SDSS DR5 and
apply these results to the physics of the BAL wind launching
mechanism.
\section{The Brightness Temperature} Equation (5) of \citet{zho06} is not well defined since it is unclear
what frame of reference is used to evaluate the quantities contained
within. Apparently, this gave rise to estimates that are $(1+z)^{3}$
larger than what we find in equation (2.4), below. In this section,
we derive a formula for $T_{b}$. Physically, it is $T_{b}$ evaluated
in the cosmological frame of reference of the quasar, $(T_{b})_{q}$,
that is relevant for assessing the "inverse Compton catastrophe." We
want to express this in terms of observable quantities at earth
designated by the subscript "o." First of all, the brightness
temperature is the equivalent blackbody temperature of the radiation
assuming one is in the Planck regime, $h \nu \ll k_{b}T$. Consider a
source in which the monochromatic intensity has increased by an
amount $I(\nu)_{q}$ in a time $\Delta t_{q}$. The brightness
temperature associated with the change in luminosity is
\begin{eqnarray}
&& (T_{b})_{q}= \frac{I(\nu)_{q} \mathrm{c}^{2}}{2 k_{b}
\nu_{q}^{2}}\;.
\end{eqnarray}
From the monochromatic version of Liouville's theorem, \citet{gun78}
\begin{eqnarray}
&& \frac{(F_{\nu})_{o}}{\Omega_{o}}\equiv I(\nu)_{o} =
\frac{1}{(1+z)^{3}} I(\nu)_{q}\;,
\end{eqnarray}
$(F_{\nu})_{o}$, is the flux density observed at earth. The solid
angle subtended by the source, $\Omega_{o}$, is bounded by the
causality requirement that the source could not have expanded more
rapidly than the speed of light during a time, $\Delta t_{q}$,
\citet{gun78},
\begin{eqnarray}
&& \Omega_{o}=\frac{\mathrm{proper\; area\; perpendicular\; to \;
line \; of \; sight}}{(\mathrm{angular\; diameter\; distance})^{2}}
\leq \frac{(c \pi \Delta t_{q})^{2}}{4 d_{A}^{2}} \;,
\end{eqnarray}
where the angular diameter distance is $d_{A}$. In a a cosmology
with $H_{0}$=70 km/s/Mpc, $\Omega_{\Lambda}=0.7$ and
$\Omega_{m}=0.3$, we can use the expression for $d_{A}$ given by
\citet{pen99}, that is accurate to $<1 \% $ relative error, along
with (2.1) -(2.3) to find:
\begin{mathletters}
\begin{eqnarray}
&& (T_{b})_{q}\approx \frac{8.0\times 10^{12}
(1+z)}{(\nu_{o}/\mathrm{1GHz})^{2}(\Delta t_{o}/ 1 \mathrm{yr})^{2}}
Z^{2} (\Delta F_{\nu}(\mathrm{mJy}))_{o}\,^{\circ}\,\mathrm{K}\;,\\
&& Z \equiv 3.31-(3.65) \nonumber \\
&&\times\left(\left[(1+z)^{4}-0.203(1+z)^{3}+0.749(1+z)^{2}
+0.444(1+z)+0.205\right]^{-0.125}\right)\;,
\end{eqnarray}
\end{mathletters}
where $\Delta F_{\nu}(\mathrm{mJy})$ is the change in flux density
in mJy measured at earth at frequency $\nu_{o}$ during the time
interval $\Delta t_{o}$.
\par When $(T_{b})_{q}> 10^{12}\,^{\circ}\,\mathrm{K}$, the inverse
Compton catastrophe occurs. Most of the electron energy is radiated
in the inverse Compton regime. The radio synchrotron spectrum from
the jet is diminished in intensity to unobservable levels
\citep{kel69}. In order to explain the observed radio synchrotron
jet in such sources, Doppler boosting is customarily invoked to
resolve the paradox. Recall that for an unresolved source the
observed flux density, $(F_{\nu})_{o}$, is Doppler enhanced relative
to the intrinsic flux density,
$(F_{\nu})_{o}=\delta^{3+\alpha}(F_{\nu})_{\mathrm{intrinsic}}$,
where the Doppler factor, $\delta$ is given in terms of $\Gamma$,
the Lorentz factor of the outflow; $\beta$, the three velocity of
the outflow; and $\alpha$, is the spectral index,
$\delta=1/[\Gamma(1-\beta\cos{\theta})]$ \citep{lin85}. Thus,
Doppler beaming can cause $(T_{b})_{q}$ to be enhanced by a factor
of $\delta^{3+\alpha}$ in (2.4). Assuming a flat spectral index for
the unresolved relativistic jet, we choose $\alpha=0$. This allows
us to find a minimum Doppler factor that avoids the inverse Compton
catastrophe, $\delta_{min}[(T_{b})_{q}]=[(T_{b})_{q}/
(10^{12}\,^{\circ}\,\mathrm{K})]^{0.333}$. If the jet plasma
propagates with $\delta > \delta_{min}[(T_{b})_{q})]$ then the
intrinsic $T_{b}$ in the frame of reference of the jet plasma is
sufficiently low that there is no inverse Compton catastrophe.
\begin{table}
\caption{Properties of the Radio Variable BAL
Quasars\tablenotemark{a}} {\footnotesize \begin{tabular}{ccccccccc}
\tableline \rule{0mm}{3mm}
 Name &   $z$ & $S^{P}_{F}$:date & $S^{P}_{N}$:date &  $\sigma$  & BI CIV & BI AlIII & BI MgII & Type \\
  (SDSS J)     &    &  (mJy) & (mJy)   &  & (km/s) &(km/s) & (km/s) & \\
\tableline \rule{0mm}{3mm}
 081102.91+500724.5\tablenotemark{b} &1.84 & 23.07: 05/23/97 & 18.80: 11/15/93 & 5.7  & 1568 & 468 & ... & LoBAL \\
 081618.99+482328.4& 3.57 & 72.29: 05/01/97 & 61.10: 11/15/93  &  5.2   & 2576 &  ... &....& HiBAL \\
 081839.00+313100.1& 2.37 & 8.82: 10/23/95 & 7.2: 12/15/93  &  3.1   & 6445 &  ... &....& HiBAL \\
 082817.25+371853.7\tablenotemark{c} &1.35 & 21.18: 07/23/94 & 14.50: 12/15/93 & 10.5  & ... & 428 & 1564 & LoBAL \\
 093348.37+313335.2& 2.60 &18.35: 10/23/95 & 15.90: 12/15/93  &  3.7   & 4107 &  1405 &....& LoBAL \\
 104106.05+144417.4&3.01&27.46: 12/99 & 19.00: 12/06/93  &  11.4   & 2527 &  ... & .... & HiBAL \\
 113445.83+431858.0 & 2.18 &27.38: 2/20/97 & 24.90: 11/15/93  & 3.4  & 10643 & ... & 1149 & LoBAL \\
 134652.72+392411.8& 2.47 &3.60: 08/19/94 & 2.20: 04/16/95 & 3.0  & 1505 & 557 & .... & LoBAL \\
 142610.59+441124.0& 2.68 & 6.78: 03/27/97 & 5.0: 03/12/95 & 3.6  & 8652 & 1389 & .... & LoBAL \\
 145926.33+493136.8 \tablenotemark{d}&2.37 & 5.22: 04/17/97 & 3.60: 03/12/95 & 3.4  & 9039 & 329 & ....& LoBAL \\
 155633.77+351757.3 &1.49 &30.92: 07/03/94 & 26.90: 04/16/95 & 4.3  & e & ... & .... & FeLoBAL \\
 165543.24+394519.9 &1.75 &10.15: 08/19/94 & 8.50: 04/16/95 & 3.0  & 5805 & ... & .... & HiBAL \\
\end{tabular}}
\tablenotetext{a}{The first column in table 1 is the source name,
followed by the redshift. Columns (3) and (4) are the FIRST and NVSS
peak flux densities at 1.4 GHz, respectively. Column (5) is the
statistical significance of the radio variability computed using
(3.1). Columns (6) - (8) are the BALnicity indices, \citet{wey91},
for CIV, AlII and MgII, respectively. The last column denotes
whether the source is a low ionization or a high ionization BALQSO.}
\tablenotetext{b}{There is a new spectrum in SDSS with much higher
S/N than the spectrum in \citet{zho06} that
 clearly shows broad Al III absorption}
\tablenotetext{c}{from \citet{zho06}, however there is now a much
higher signal to noise spectrum in SDSS than the one used in
\citet{zho06} and we have recomputed the BALnicity indices}
\tablenotetext{d}{from \citet{zho06}} \tablenotetext{e}{identified
as FeLoBALQSO in \citet{bec97}}
\end{table}
\section{Polar BALQSOs} Table 1 is a list of BALQSOs in which the
radio variability requires that the jet is propagating well within
$35\,^{\circ}$ to the line of sight in order for the jet plasma to
satisfy $(T_{b})_{q} < 10^{12}\,^{\circ}\,\mathrm{K}$. Since rather
small variations of flux density create these conditions we must
first prove that the sources are truly variable. We choose the
condition for variability to be
\begin{equation}
\sigma_{var}=\frac{S_{FP}-S_{NP}}{\sqrt{\sigma^2_{FP}+\sigma^2_{NP}}}>+3\;,
\end{equation}
where $S_{FP}$ and $S_{NP}$ denote the peak flux density at 1.4 GHz
measured by the FIRST and NVSS surveys, respectively, $\sigma_{FP}$
and $\sigma_{NP}$ are the FIRST and NVSS peak flux uncertainties,
respectively. The beam for the NVSS survey is considerably larger
than FIRST ($FWHM\sim 45^{''}$ versus $FWHM\sim 5^{''}$). Thus, a
compact source $\ll 5^{''}$ (like a compact radio core which is the
putative site of variable activity) should be detected with equal
sensitivity if it exceeds the flux limit of the samples. The NVSS
beam should pick up more extended flux than FIRST. Variable BALQSO
fields in NVSS with confusion from nearby sources were omitted from
table 1. Thus, a FIRST peak flux density larger than an NVSS peak
flux density by $3\sigma$ (\textbf{hence the plus sign on the RHS of
eqn. (3.1)}) is true variability and not an artifact of the
different beam sizes. One improvement from \citet{zho06} is that we
used the peak NVSS flux density as opposed to the integrated NVSS
flux density. This is preferred for the comparison described in
(3.1), since the variable radio flux is most likely coming from
sub-arcsecond regions of the jet and this choice is less sensitive
to low surface brightness noise that can be picked up in the
integrated flux \citep{con97}.
\par It is crucial to assess the significance of the $3 \sigma$
results in table 1. In order to clarify the statistics, we must
delineate the super sample from which the sources in table 1 are
drawn. The potential variable BALQSOs in DR5 that can be detected by
equation (3.1) require two conditions. First of all, the sources
must have a BALnicity index $> 0$ as defined in \citet{wey91} for at
least one line either, CIV, AlIII, or MgII. Secondly, $ S_{FP} >
3.55$ mJy. The NVSS survey has a peak flux limit of 2.5 mJy in
general, but occasionally as low as 2.2 mJy and a minimum
measurement uncertainty of $\approx 0.45$ mJy \citet{con98}. Thus, $
S_{FP} < 3.55$ mJy could never satisfy eqn. (3.1). There are 116
BALQSOs that satisfy these criteria and they form the master sample.
Eqn. (3.1) is a one sided probability, so if the errors are Gaussian
distributed as in \citet{con97,con98} then there is less than a 20\%
chance of expecting a variable BALQSO from this sample of 116.
However, we found 20 variable BALQSOs (12 of which made the table
because $T_{B}$ was sufficiently high to restrict the line of sight
to $< 35^{\circ}$). Thus, based on Gaussian statistics our findings
are very significant. However, if the errors in the tail of the
distribution are not Gaussian, but arise from unforseen sources of
measurement error, then the analysis of \citet{con97,con98} does not
apply and we have no means to assess the significance of our
results. Note that there are 4 sources in table 1 with a variability
that is significant above the $5\sigma$ level.

\begin{table} \caption{Line of Sight to Radio Variable BAL
Quasars} {\footnotesize
\begin{tabular}{ccccc} \tableline \rule{0mm}{3mm}
 Name &   $T_{b}$ & $\delta_{\mathrm{min}}$ & $\theta_{\mathrm{max}}$ & Type \\
  (SDSS J)     & ($10^{12}$$\,^{\circ}$ K)   &   & (degrees)   &   \\
\tableline \rule{0mm}{3mm}
 081102.91+500724.5 & 5.4 & 1.7 & 34.9 & LoBAL \\
 081618.99+482328.4& 39.6 & 3.4 & 17.1  & HiBAL \\
 081839.00+313100.1& 11.8 & 2.3 & 26.1  & HiBAL \\
 082817.25+371853.7& 125.4 & 5.0 & 11.5  & LoBAL \\
 093348.37+313335.2& 20.2 & 2.7 & 21.6  & LoBAL \\
 104106.05+144417.4& 8.5 & 2.0 & 29.4 & HiBAL \\
 113445.83+431858.0 & 5.9 & 1.9 & 33.7 & LoBAL \\
 134652.72+392411.8& 73.9 & 4.2 & 13.8 & LoBAL \\
 142610.59+441124.0& 11.5 & 2.3 & 26.4 & LoBAL \\
 145926.33+493136.8 & 8.8 & 2.1 & 29.0 & LoBAL \\
 155633.77+351757.3 & 66.5 & 4.0 & 14.3 & FeLoBAL \\
 165543.24+394519.9 & 47.5 & 3.6 & 16.1 & HiBAL \\

\end{tabular}}

\end{table}
\par Our BALQSO identifications and the BALnicity indices in table 1 were derived from spectra in the SDSS DR5
public archive. The SDSS spectra of radio-variable BALs were
retrieved from the SDSS database and were analyzed using the
$IRAF$\footnote{IRAF is the Image Reduction and Analysis Facility,
written and supported by the IRAF programming group at the National
Optical Astronomy Observatories (NOAO) in Tucson, Arizona.}
software. First, the spectrum was de-reddened using the Galactic
extinction curve, \citet{sch98}, then the wavelength scale was
transformed from the observed to the source frame. The spectra were
fitted in XSPEC with a powerlaw plus multiple gaussian model,
including fits for both emission lines and absorption lines
\citep{arn96}. All the model parameters were kept free. The best fit
to the SDSS data was determined using $\chi ^{2}$ minimization. If
an emission bump around 2500 $\AA$ is present then the continuum fit
in the CIV (1550 $\AA$) region was extrapolated to longer
wavelength. The BALnicity indices quoted in table 1 are a
consequence of the method of spectral fitting described above and
other methods might produce different results. However, the exact
BALnicity index is not critical to this discussion, the sources in
table 1 clearly show BALs and that is the essential point of
relevance here.
\par In table 2, we calculate $(T_{b})_{q}$, using equation (2.4)
and the peak fluxes from table 1. From this, we calculate
$\delta_{min}[(T_{b})_{q}]$ in column (3) as described at the end of
section 2. For each value of $\delta_{min}$, one can vary $\beta$ in
the definition of $\delta$ to find the maximum value of $\theta$,
$\theta_{max}\{\delta_{min}[(T_{b})_{q}]\}$, that is compatible with
$\delta_{min}$:
\begin{equation}
\theta_{max}\{\delta_{min}[(T_{b})_{q}]\}=\mathrm{Max}_{\mid_\beta}\left(\arccos\left\{\left[1-\left(\frac{\sqrt{1-\beta^{2}}}{\delta_{min}[(T_{b})_{q}]}\right)\right]\beta^{-1}\right\}\right)\;.
\end{equation}
The value of $\theta_{max}$ in column (4) is an extreme upper limit
because it is the third in a chain of bounds. \begin{itemize}
\item The value of $(T_{b})_{q}$ in table 1 is a lower bound, since
NVSS might pick up some extended flux that is missed by the peak
FIRST measurement and the radio core was actually weaker at the time
of the NVSS measurement than indicated by the peak flux density,
i.e., $(\Delta F_{\nu}(\mathrm{mJy}))_{o}$ is underestimated in
(2.4). Also, (2.3) is an inequality. From the discussion at the end
of section 2, the larger $(T_{b})_{q}$, the larger $\delta_{min}$
\item the larger $\delta_{min}$, the lower $\theta_{max}$ by (3.2)
and the true value of the jet plasma $\delta$ can be much larger
than $\delta_{min}[(T_{b})_{q}]$
\item the larger $\delta_{min}$, the lower $\theta_{max}$ by (3.2)
and the true value of the jet plasma $\delta$ can be much larger
than $\delta_{min}[(T_{b})_{q}]$
\item The actual line of sight to earth satisfies
$\theta<\theta_{max}$ by definition.
\end{itemize} Thus the condition for inclusion
into table 2, $\theta_{max}< 35\,^{\circ}$, means that the jet is
likely to be propagating very close to the pole.

\section{The Theory of Polar BALQSOs}
Figure 1 is a small modification to Fig. 9 of \citet{pun00}. In
section 4.5 of \citet{pun00}, it is discussed how a relativistic jet
(in red) can coexist nested inside of the bipolar BAL wind. This jet
can emit radio flux that is beamed to within a small angle of the
polar axis. Within the axisymmetric version of the model, LoBALQSOs
exist for lines of sight within $15^{\circ}$ of the polar axis. The
density of the wind is highest nearest the polar axis as indicated
by the gray shading in figure 1 and LoBALQSOs are viewed closer to
the polar axis on average than HiBALQSOs. The accretion disk coronal
X-rays are screened from the BAL wind by the dense base of the jet
which provides hydrogen column densities of $\Sigma_{H}\sim
10^{25}\mathrm{cm^{-2}}$. Similarly, the different lines of sight
near the polar axis, through the inhomogeneous wind, naturally
provide larger Compton scattering columns and more attenuation of
the far UV flux from the inner disk and less attenuation of the near
UV and optical flux from the outer disk, making the spectrum appear
red. The more extreme and highly polarized LoBALs obviously occur in
objects without a perfectly axisymmetric distribution of dust as
discussed in detail in \citet{pun00}. Recall that the bipolar BAL
wind does not preclude the coexistence of a BAL wind from the outer
regions of the accretion disk as envisioned by \citet{mur95}.
\par Since the LoBALQSOs are viewed closer to
the jet axis than the HiBALQSOs in the axisymmetric model, geometric
arguments imply they should typically have jets with larger Doppler
factors. Thus, LoBALQSOs should occur in samples of variable BALQSOs
at an inordinately high rate if the axisymmetric version of the
polar model applies to a significant subpopulation of the BALQSOs.
In our master sample of 116 FIRST, DR5 BALQSOs defined in section 3,
69 are HiBALQSOs and 47 are LoBALQSOs. Based on table 1, 4/69 =
5.80\% of HiBALQSOs and 8/47 = 17.02\% of the LoBALQSOs have
$(T_{b})_{q}> 5\times 10^{12}\,^{\circ}\,\mathrm{K}$ (this condition
is equivalent to $\theta < 35^{\circ}$). The sample is small, but
the fact that in table 1, the LoBALQSO likelihood to have a large
$T_{b}$ is 2.94 times that of the HiBALQSOs tends to support the
notion the polar BALQSO model represents a significant subpopulation
of BALQSOs.
\section{Conclusion} Using radio variability arguments,
we expanded on the sample of known polar BALQSOs begun by
\citet{zho06}. In the process, we noted that these radio variable
BALQSOs have an inordinately large LoBALQSOs subpopulation. It is
interesting that these properties are expected based on an existing
detailed theoretical treatment and modeling of bipolar BAL winds
\citep{pun99,pun00}. It would be informative to continue to monitor
the FIRST BALQSOs at 1.4 GHz, with matched resolution, in order to
find more $(T_{b})_{q}> 10^{12}\,^{\circ}\,\mathrm{K}$ BALQSOs and
improve the statistical information in table 2.

\clearpage

\begin{figure}
    \begin{center}
        \includegraphics[scale=1.0]{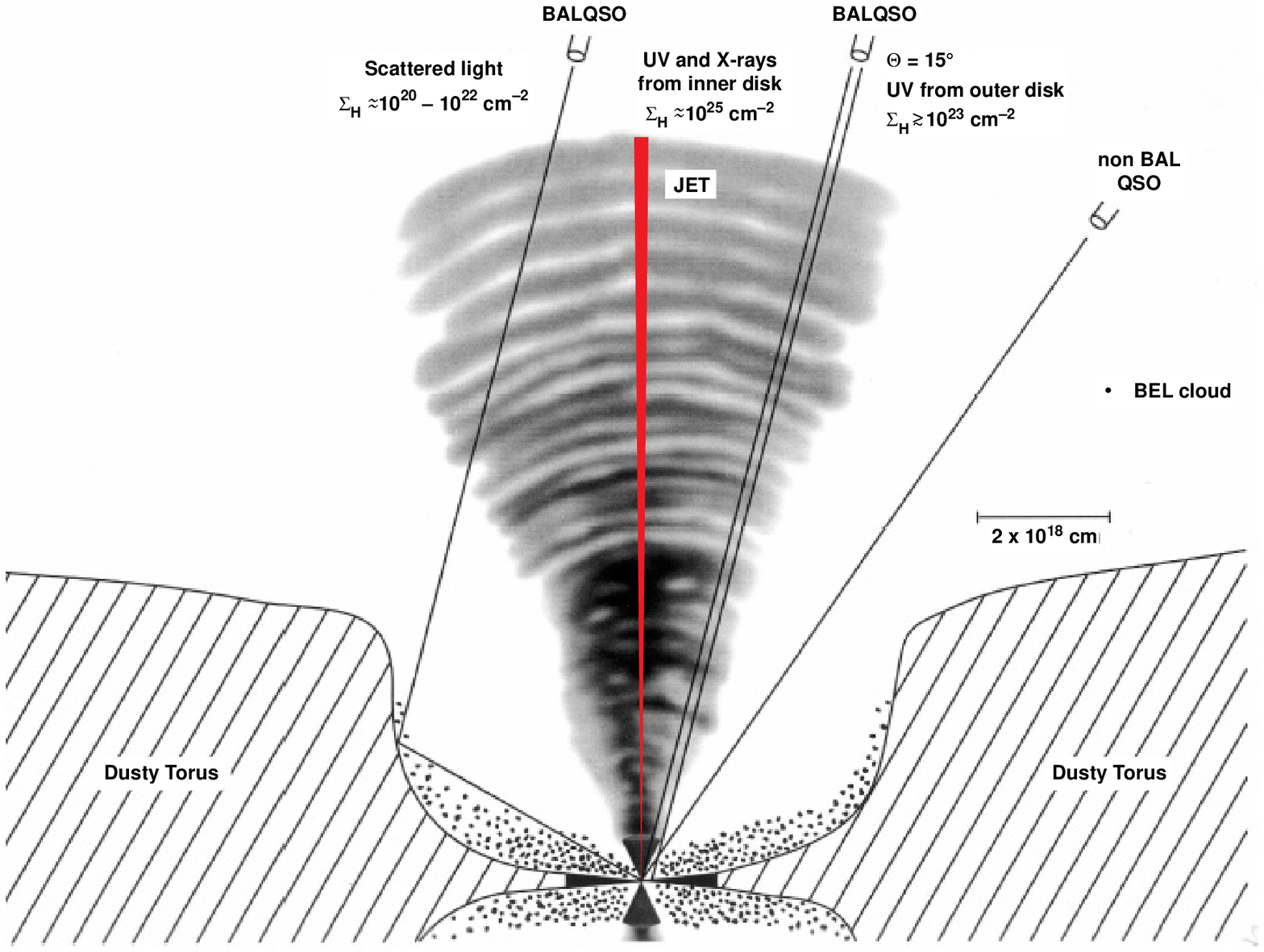}
    \end{center}
    \caption{The bipolar wind model of \citet{pun99,pun00}. The gray clouds of the BAL wind has an opening angle of $\sim 25^{\circ}$ - $30^{\circ}$.
    Nested inside is a relativistic jet (in red). The highest BAL wind densities are near the polar axis. Thus, lines of sight
    near the polar axis, have the maximum attenuation of X-ray and ionizing UV
radiation from the accretion flow. This implies that lines of sight
close to the polar axis are more likely to represent LoBALQSOs.}
    \end{figure}

\clearpage

\end{document}